\documentclass[aps,prl,reprint,superscriptaddress,twocolumn,10pt]{revtex4-2}

\usepackage{graphicx}% Include figure files
\usepackage{dcolumn}% Align table columns on decimal point
\usepackage{bm}% bold math
\usepackage{hyperref}% add hypertext capabilities
\usepackage{amsmath}
\usepackage{xcolor}
\usepackage{comment}

\begin{document}

%\preprint{}

\title{Exploring the Onset of Collectivity Approaching N=40 through Manganese Masses}
% \title{Probing the N=40 Island of Inversion through Manganese Masses}


\author{C. Chambers}
\email[Corresponding author: ]{cchambers2@triumf.ca}
    \affiliation{TRIUMF, 4004 Wesbrook Mall, Vancouver, British Columbia V6T 2A3, Canada}
    
\author{M.P. Reiter}
    %\email[Corresponding author: ]{mreiter@ed.ac.uk}
    \affiliation{II. Physikalisches Institut, Justus-Liebig-Universit\"{a}t, 35392 Gie{\ss}en, Germany}  
    \affiliation{TRIUMF, 4004 Wesbrook Mall, Vancouver, British Columbia V6T 2A3, Canada}
    \affiliation{School of Physics and Astronomy, University of Edinburgh, Edinburgh EH9 3FD, UK}

\author{A. T. Gallant}  
    \affiliation{Physicial Life Sciences Directorate, Lawrence Livermore National Laboratory, Livermore, CA 94550, USA}

\author{M. Yavor}
    \email[Results presented here were obtained before 2022.]{}
    \affiliation{II. Physikalisches Institut, Justus-Liebig-Universit\"{a}t, 35392 Gie{\ss}en, Germany} 
    \affiliation{Institute for Analytical Instrumentation, Russian Academy of Sciences, 190103 St. Petersburg, Russia}

\author{C. Andreoiu}
    \affiliation{Department of Chemistry, Simon Fraser University, Burnaby, BC V5A 1S6, Canada}
    \affiliation{Department of Physics, Simon Fraser University, Burnaby, BC V5A 1S6, Canada}

\author{C. Babcock}
    \affiliation{TRIUMF, 4004 Wesbrook Mall, Vancouver, British Columbia V6T 2A3, Canada}

\author{J. Bergmann}
    \affiliation{II. Physikalisches Institut, Justus-Liebig-Universit\"{a}t, 35392 Gie{\ss}en, Germany}     

\author{T. Dickel}
    \affiliation{II. Physikalisches Institut, Justus-Liebig-Universit\"{a}t, 35392 Gie{\ss}en, Germany}  
    \affiliation{GSI Helmholtzzentrum f\"{u}r Schwerionenforschung GmbH, 64291 Darmstadt, Germany}

\author{J. Dilling}
    \affiliation{TRIUMF, 4004 Wesbrook Mall, Vancouver, BC V6T 2A3, Canada}
    \affiliation{Department of Physics \& Astronomy, University of British Columbia, Vancouver, BC V6T 1Z1, Canada}
    \affiliation{Oak Ridge National Laboratory, Oak Ridge, TN 37830, USA}

%\author{I. Dillmann}
%	\affiliation{TRIUMF, 4004 Wesbrook Mall, Vancouver, British Columbia V6T 2A3, Canada}
%    \affiliation{Department of Physics and Astronomy, University of Victoria, Victoria, British Columbia V8P 5C2, Canada}    

\author{E. Dunling}
    \affiliation{TRIUMF, 4004 Wesbrook Mall, Vancouver, British Columbia V6T 2A3, Canada}
    \affiliation{Department of Physics, University of York, York, YO10 5DD, United Kingdom}

\author{G. Gwinner}
    \affiliation{Department of Physics \& Astronomy, University of Manitoba, Winnipeg, MB R3T 2N2, Canada}   

\author{Z. Hockenbery}
    \affiliation{TRIUMF, 4004 Wesbrook Mall, Vancouver, British Columbia V6T 2A3, Canada}
    \affiliation{Department of Physics, McGill University, Montreal, QC H3A 2T8, Canada }  

\author{J.D. Holt}
    \affiliation{TRIUMF, 4004 Wesbrook Mall, Vancouver, British Columbia V6T 2A3, Canada}
    \affiliation{Department of Physics, McGill University, Montreal, QC H3A 2T8, Canada }

\author{R. Klawitter}
    \affiliation{TRIUMF, 4004 Wesbrook Mall, Vancouver, British Columbia V6T 2A3, Canada}
    \affiliation{Max-Planck-Institut f\"{u}r Kernphysik, Heidelberg D-69117, Germany}
		
\author{B. Kootte}
    \affiliation{TRIUMF, 4004 Wesbrook Mall, Vancouver, British Columbia V6T 2A3, Canada}
    \affiliation{Department of Physics \& Astronomy, University of Manitoba, Winnipeg, MB R3T 2N2, Canada}
    \affiliation{Department of Physics, University of Jyväskylä, 40014 Jyväskylä, Finland}

\author{Y. Lan}
    \affiliation{TRIUMF, 4004 Wesbrook Mall, Vancouver, British Columbia V6T 2A3, Canada}
    \affiliation{Department of Physics \& Astronomy, University of British Columbia, Vancouver, BC V6T 1Z1, Canada}

\author{J. Lassen}
    \affiliation{TRIUMF, 4004 Wesbrook Mall, Vancouver, British Columbia V6T 2A3, Canada}

\author{E. Leistenschneider}
    \affiliation{TRIUMF, 4004 Wesbrook Mall, Vancouver, British Columbia V6T 2A3, Canada}
    \affiliation{Department of Physics \& Astronomy, University of British Columbia, Vancouver, BC V6T 1Z1, Canada}
    \affiliation{Lawrence Berkeley National Laboratory, Berkeley, CA 94720, USA}

\author{R. Li}
    \affiliation{TRIUMF, 4004 Wesbrook Mall, Vancouver, British Columbia V6T 2A3, Canada}
    \affiliation{Department of Physics, University of Windsor, Windsor, ON N9B 3P4, Canada}
  	
\author{T. Miyagi}
    \affiliation{TRIUMF, 4004 Wesbrook Mall, Vancouver, British Columbia V6T 2A3, Canada}
    \affiliation{Center for Computational Sciences, University of Tsukuba, Ibaraki-ken 305-8577, Japan}

\author{M. Mostamand}
    \affiliation{TRIUMF, 4004 Wesbrook Mall, Vancouver, British Columbia V6T 2A3, Canada}
    \affiliation{Department of Physics \& Astronomy, University of Manitoba, Winnipeg, MB R3T 2N2, Canada}
    \affiliation{Paul Scherrer Institute PSI, 5232 Villigen PSI, Switzerland}

\author{W.R. Pla\ss}
    \affiliation{II. Physikalisches Institut, Justus-Liebig-Universit\"{a}t, 35392 Gie{\ss}en, Germany}  
    \affiliation{GSI Helmholtzzentrum f\"{u}r Schwerionenforschung GmbH, Planckstra{\ss}e 1, 64291 Darmstadt, Germany}    

\author{C. Scheidenberger}
    \affiliation{II. Physikalisches Institut, Justus-Liebig-Universit\"{a}t, 35392 Gie{\ss}en, Germany}  
    \affiliation{GSI Helmholtzzentrum f\"{u}r Schwerionenforschung GmbH, Planckstra{\ss}e 1, 64291 Darmstadt, Germany} 

\author{R. Thompson}
    \affiliation{Department of Physics and Astronomy, University of Calgary, Calgary, AB T2N 1N4, Canada}

\author{M. Vansteenkiste}
    \affiliation{TRIUMF, 4004 Wesbrook Mall, Vancouver, British Columbia V6T 2A3, Canada}
    \affiliation{Department of Physics and Astronomy, University of Waterloo, Waterloo, ON N2L 3G1, Canada}
    \affiliation{Department of Physics, University of Toronto, Toronto, ON M5S 1A7, Canada}

\author{M.E. Wieser}
    \affiliation{Department of Physics and Astronomy, University of Calgary, Calgary, AB T2N 1N4, Canada}   

\author{A.A. Kwiatkowski}
    \affiliation{TRIUMF, 4004 Wesbrook Mall, Vancouver, British Columbia V6T 2A3, Canada}
    \affiliation{Department of Physics and Astronomy, University of Victoria, Victoria, BC V8P 5C2, Canada}


\date{\today}

% --------------------- ABSTRACT --------------
\begin{abstract}
Isotopes in the region of the nuclear chart below $^{68}\mathrm{Ni}$ have been the subject of intense experimental and theoretical effort due to the potential onset of a new ``island of inversion'' when crossing the harmonic oscillator subshell closure at $N = 40$.
We have measured the masses of $^{64-68}\textrm{Mn}$ using TITAN's multiple-reflection time-of-flight mass spectrometer, resulting in the first precision mass measurements of $^{67}\mathrm{Mn}$ and $^{68}\mathrm{Mn}$. These results are compared to \textit{ab initio} calculations and modern shell model calculations and show an increase in collectivity approaching $N=40$.
\end{abstract}

%\keywords{}

\maketitle

% ----------------------------------------------------
% ------------------- Introduction -------------------
% ----------------------------------------------------

The region around $(Z,N) \approx (28,40)$ has been a focus of much experimental effort due to observation of the large $E(2^+_1)$ value in $^{68}\mathrm{Ni}$ \cite{BERNAS1982279,PhysRevLett.74.868}, which had been initially interpreted as a $N=40$ neutron-subshell closure.
However, it has been argued that the large energy difference between the $0^{+}$ ground state and $2^{+}$ excited state in $^{68}\mathrm{Ni}$ is the result of mixing of two underlying $0^{+}$ configurations \cite{RevModPhys.83.1467,PhysRevC.82.027304}.
In particular, it seems to be the same underlying physics at play in $^{68}\mathrm{Ni}$ as in $^{90}\mathrm{Zr}$, mainly, the interplay between closed shells at $Z = 28$ and $Z = 50$ and the harmonic oscillator subshell closure at $Z,N = 40$.
This behavior has been explained as the excitation of protons across the $Z = 28$ shell closure and the resulting gain in correlation energy due to pairing correlations and residual proton-neutron interactions \cite{RevModPhys.83.1467}.
The systematics in this area point towards an onset of collectivity towards $N = 40$ \cite{PhysRevC.86.014325}.
For example, given both the systematics of the mass surface in the nickel isotopes and the relatively large $B(E2)$ values, one expects a sharp drop at $^{68}\mathrm{Ni}$, which only supports a relatively weak subshell closure.

Measurements in the manganese isotopes further point to the onset of collectivity towards $N = 40$.
Co-linear laser spectroscopy measurements of the $g$-factor show a decrease towards $N=40$ \cite{PhysRevC.92.044311}, while measurements of the charge radius increase towards $N=40$ \cite{PhysRevC.94.054321}. Both are indicative of the onset of deformation.
Shell model calculations in this region show a parallel increase in the neutron occupancies towards $N = 40$ and in the proton occupancies across the $N = 28$ shell gap \cite{PhysRevC.94.054321}.
Recently, the excited states of the odd-even isotopes $^{63,65,67}\mathrm{Mn}$ have been studied, and show strongly-coupled rotational bands that are to be expected in deformed odd $A$ nuclei \cite{LIU2018392}.
This points to the sensitive and complex interplay between the protons in the $pf$ shell and the neutrons in the $pf$ and $gd$ shells.

Above the Mn isotopes, in the Fe isotopic chain, measurements of the reduced $E2$ transition probabilities $B(E2;2^+ \rightarrow 0^+)$ in the even-even isotopes clearly show a region of enhanced collectivity at $N = 40$ \cite{PhysRevLett.106.022502}.
The two-neutron separation energies do not show any indication of a subshell closure beyond $N = 40$ \cite{PhysRevC.81.044318}.
One-proton knockout reactions \cite{PhysRevC.92.034306} in the Cr isotopes approaching $N = 40$ have shown a transition from a spherical vibrator to a deformed rigid rotor behavior in the low lying yrast states.
Recent mass measurements in the Fe and Co isotopic chains have very smooth two-neutron separation energies past $N = 40$ \cite{PhysRevC.97.014309,PhysRevC.101.041304,PhysRevC.105.L041301}, while the mass surface in both the Mn and Cr isotopic chains show an increasing trend towards collectivity as indicated by the flattening of the two-neutron separation energies towards $N = 40$ \cite{PhysRevLett.120.232501,PhysLettB.83.137288}.
This trend in the Cr mass surface is reminiscent of the behavior of the Mg isotopic chain through the island of inversion \cite{PhysRevC.88.054317,PhysRevC.101.041304}.

Here we present the results of precision mass measurements of $^{64-68}\textrm{Mn}$ with TITAN's MR-TOF-MS, extending the mass surface in this region to $N = 43$, and performing the first precision mass measurement of $^{67}\textrm{Mn}$ and $^{68}\textrm{Mn}$.
These results are compared to state-of-the-art \textit{ab initio} shell model calculations, and implications for the subshell gap at $N = 40$ are discussed.

% ----------------------------------------------------
% -------------------- Experiment --------------------
% ----------------------------------------------------

\section{Experimental Description and Results}

\begin{figure}[t]
	\includegraphics[width=0.9\linewidth]{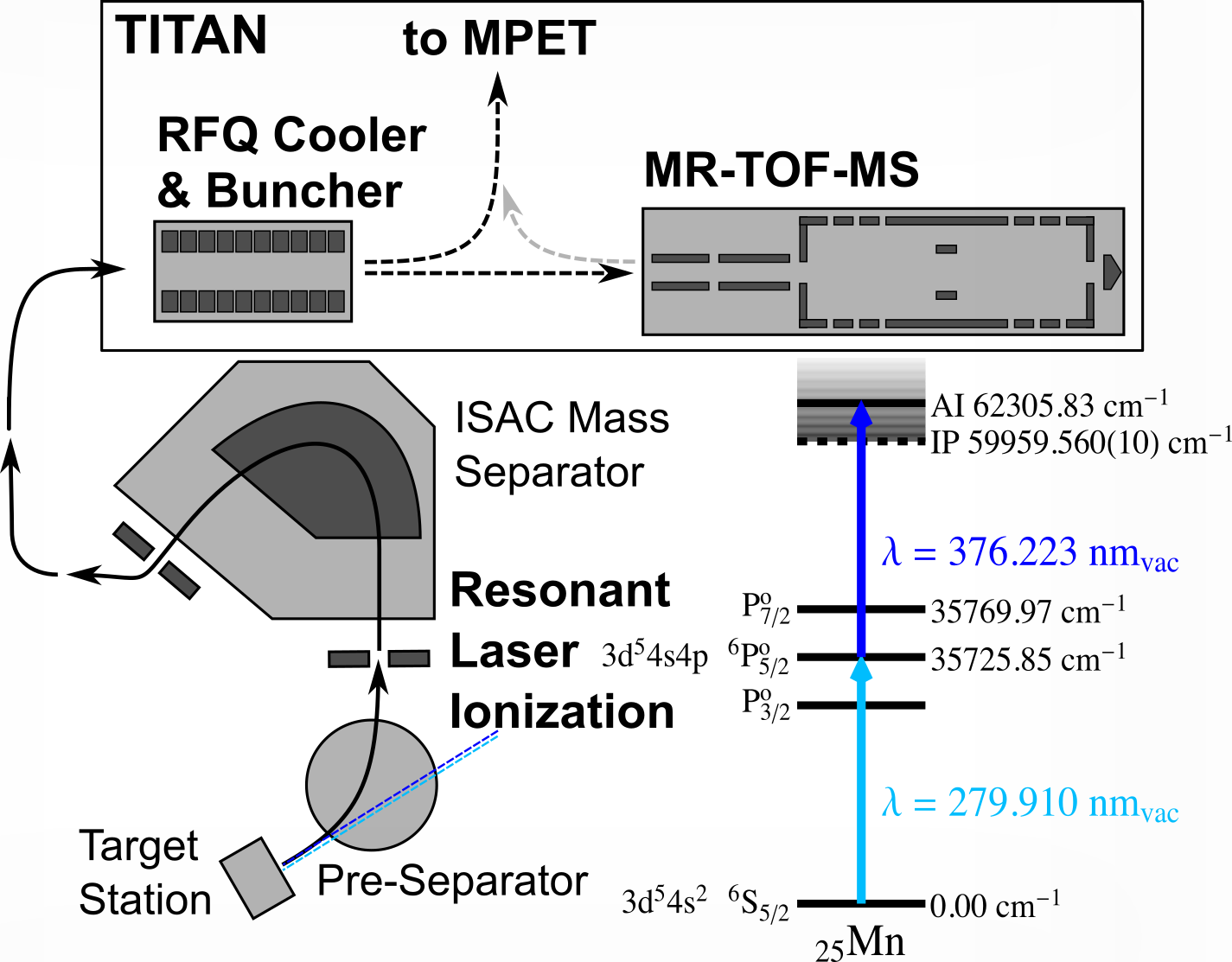}
	\caption{
		Overview of the ISAC and TITAN facilies.
		The two-step resonant laser ionization scheme used for Mn utilizes an energetically lower autoioinizing (AI) state than referenced in \cite{SALOMAN1994251,doi:10.1063/1.4850695}.
	}
	\label{Fig:TitanLayout}
\end{figure}

The mass measurements were performed using the Multiple-Reflection Time-of-Flight Mass Spectrometer (MR-TOF-MS) device \cite{Jesch2015,Dickel:2019aa,REITER2021} of TRIUMF's Ion Trap for Atomic and Nuclear science (TITAN) shown in Fig. \ref{Fig:TitanLayout}.
The beams of Mn were produced at TRIUMF's Isotope Separator and Accelerator (ISAC) facility \cite{doi:10.1063/1.1150364} by bombarding a $\mathrm{UC}_x$ target with a $480\,\textrm{MeV}$, $9.8\,\mu \textrm{A}$ proton beam.
After extraction from the target, the beams of Mn isotopes were resonantly laser ionized using TRIUMF's Resonant Ionization Laser Ion Source (TRILIS) \cite{doi:10.1063/1.3115616}.
The resonance ionization scheme for Mn used a second excitation step into an autoioinizing (AI) state of lower excitation energy than referenced in the literature \cite{SALOMAN1994251,doi:10.1063/1.4850695} as shown in Fig. \ref{Fig:TitanLayout}.
Both excitation steps were driven in saturation, with measured saturation powers of 30\,mW and 260\,mW for the first and second excitation step respectively.
During the experiment lasers were operated at about 100\,mW and 350\,mW output power.
All laser powers were measured on the laser table with a $50\%$ laser beam transport efficiency into the 3\,mm diameter ionization volume of the ISAC hot cavity target ion source.
Details of the laser system and laser ion source are given in \cite{Lassen2017} and references therein.

Along with the desired Mn beams, surface ionized contaminant species of Ni$^{+}$, Co$^{+}$, Fe$^{+}$, Zn$^{+}$, $\textrm{Ba}^{2+}$, and various diatomic oxide molecules and fluoride molecules were also delivered.
Typically, the TOF spectra were dominated by oxides formed with $^{16}\textrm{O}$ and the highest abundant stable isobar 16 mass units below the $A$ of interest, and $\mathrm{Ba}^{2+}$.
These contaminant species beams served both as calibrations and also as cross-checks on the stability and accuracy of the MR-TOF-MS measurements.

\begin{figure}[t]
	\includegraphics[width=\linewidth]{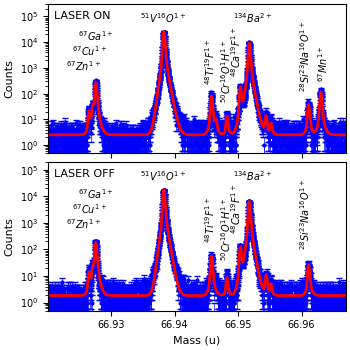}
	\caption{
		(color online)
		Time-of-flight spectra for $A = 67$ showing the large amount of contamination from the surface ionized beams of $^{51}\textrm{V}^{16}\textrm{O}^{+}$ and $^{134}\textrm{Ba}^{2+}$.
		The red lines are fits to the full spectrum of peaks with hyper-EMG lineshapes \cite{IJMS.421.245,Paul.emgfit}.
	}
	\label{Fig:A67TOF}
\end{figure}

The extracted beams were delivered continuously to TITAN's Radiofrequency Quadrupole (RFQ) cooler and buncher \cite{Brunner201232}, where collisions with a helium buffer gas cooled the beams.
The cold ion bunches were sent to the MR-TOF-MS.
A dedicated RFQ ion trap in the MR-TOF-MS collected and prepared the beam which was further cooled for $\approx 13\textrm{ms}$.
After cooling, the ions were injected into the mass analyzer for either 500 or 595 isochronous turns and one time-focusing shift turn.
The time-focusing shift turn creates a TOF focus of the ion bunches on the MCP \cite{DICKEL20171}.
This produces peak widths of approximately $23\,\textrm{ns}$ for time-of-flights of $\approx 9.7\,\textrm{ms}$, resulting in resolving powers of $\approx 210\,000$.
Clear, unambiguous identification of the $\mathrm{Mn}$ peak was accomplished by turning the ionization lasers on and off, as demonstrated in Fig. \ref{Fig:A67TOF} for $^{67}\mathrm{Mn}$. 

The atomic masses $M_{a}$ can be determined from the fitted peak positions using $M_{a} = q(C(t_{\mathrm{ion}} - t_0)^2 + m_e)$, where $m_e$ is the mass of an electron, $q$ is the charge state of the ion, $t_{\mathrm{ion}}$ is the fitted centroid of the ion's time-of-flight, $C$ is a device dependent calibration, and $t_0$ is a small offset to account for signal delay and propagation.
The time offset $t_0$ is a constant, independent of the ions being measured, and was determined from a single turn spectrum using $^{39}\textrm{K}^+$ and $^{41}\textrm{K}^+$.
Systematic contributions, such as device dependent systematics related to ringing from non-ideal electrode switching, time-dependent drifts, etc., were determined to be $\Delta M / M \approx 3 \times 10^{-7}$ following the description in \cite{PhysRevC.99.064313}. This uncertainty has since been improved upon and is now below $\Delta M / M \approx 1 \times 10^{-7}$ for most measurements, as reported in \cite{REITER2021}. In order to reduce ion-ion dependent shifts the measurements of $^{64-65}\mathrm{Mn}$ were performed with on average less than one ion per MR-TOF-MS cycle. However, due to much stronger isobaric background, the measurements of $^{66,67,68}\mathrm{Mn}$ had to be performed at higher rates and systematic shifts arising from ion-ion interaction had to be considered.

\subsection{Ion-Ion Interaction Considerations}
At high intensities, MR-TOF-MS operation is limited due to ion-ion interaction \cite{10.1063/1.4796061,PhysRevLett.87.055001,PhysRevA.65.042704} at times even fully preventing separation of different ion species, effects often refereed to as self-bunching or peak coalescence \cite{DStrasser_2003,Grinfeld2014,Hipple-ion-ion,schweikhard_2014,Rosenbusch_ion-ion,MAIER2023168545}. However, already at more moderate intensities space charge effects start to effect high precision mass measurements causing systematic shifts \cite{https://doi.org/10.1002/jms.5006,koslov-ion-ion,DICKEL2015172}.    
%lippertphd2016,

\begin{figure}[t]
	\includegraphics[width=0.9\linewidth]{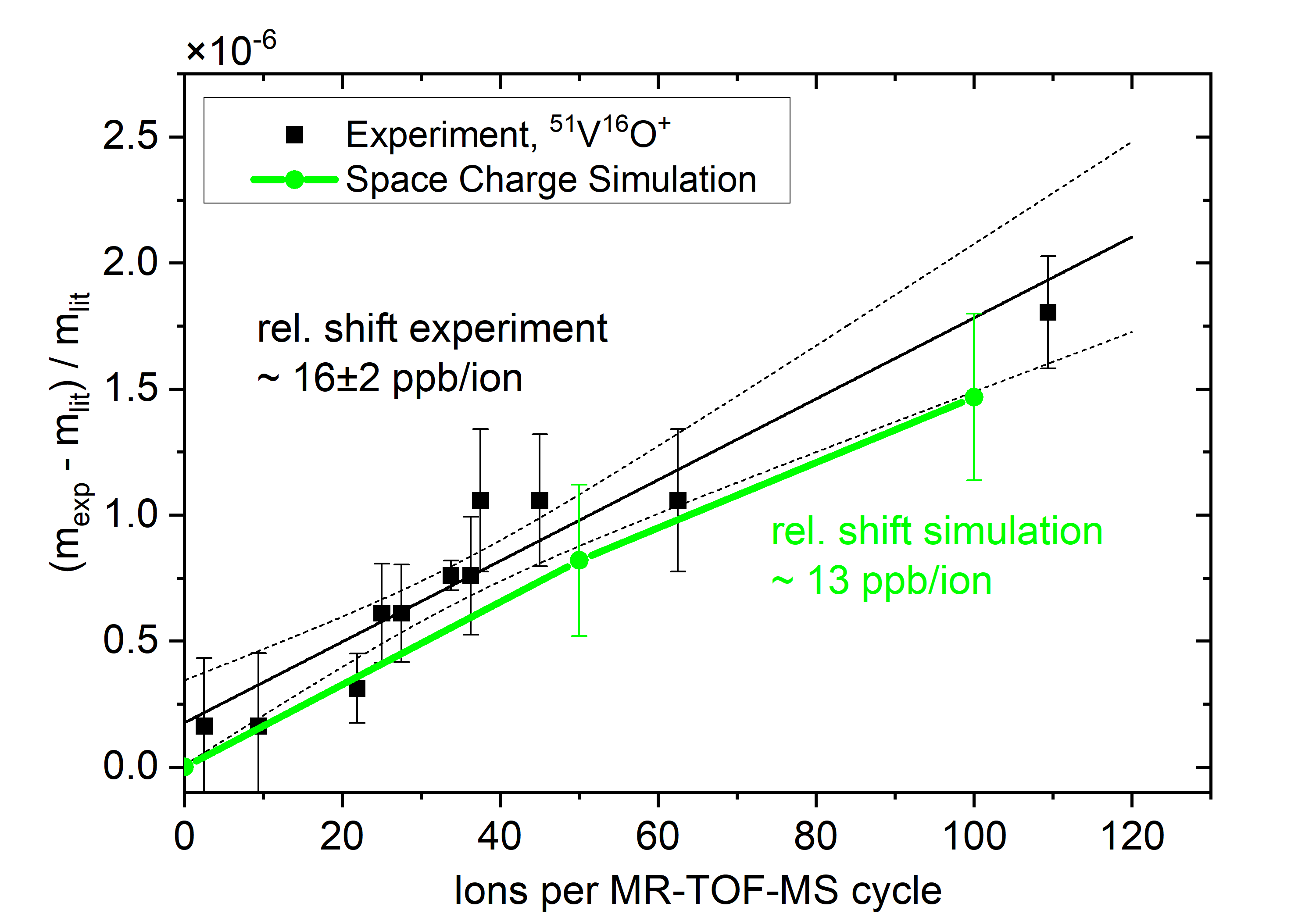}
	\caption{
		(color online)
		Mass of $^{51}\textrm{V}^{16}\textrm{O}^{+}$ based on $^{67}$Ga$^{+}$ calibration obtained at different intensities. A clear linear shift can be seen.  
	}
	\label{Fig:space_charge}
\end{figure}

To estimate the magnitude of those effects on the mass accuracy of the TITAN MR-TOF system a set of short mass measurements of $^{51}\mathrm{V}^{16}\mathrm{O}^{+}$ and $^{134}\textrm{Ba}^{2+}$ were performed at different overall ion beam intensities. 
The measured rate was corrected for both the transmission efficiency from zero to $500$ turns of about $40\%$ during this experiment and the detection efficiency of the MCP detector itself of about $40\%$ to obtain the total number of simultaneously flying ions. 
Measurements were performed up to about $100$ simultaneously flying ions per MR-TOF-MS cycle at which point the MCP detector and DAQ systems started to be affected by dead-time effects. The given rate per cycle was corrected for dead-time effects based on the known single-ion event width of $1.6$~ns. The later, was estimated from a simultaneous measurement of $^{39}$K and $^{41}$K ions at different intensities.   
Each mass spectrum was analyzed according to the previous procedure and the masses of $^{51}\mathrm{V}^{16}\mathrm{O}^{+}$ and $^{134}\textrm{Ba}^{2+}$ based on a $^{67}$Ga$^{+}$ calibration were obtained.
Fig.~\ref{Fig:space_charge} shows the variation of the mass of $^{51}\mathrm{V}^{16}\mathrm{O}^{+}$.
A clear shift of the mass value even up to $\approx 2$~ppm can be seen as function of beam intensity.  
Based on a linear fit we obtained a systematic shift of $16\pm2$~ppb per ion for $^{51}\mathrm{V}^{16}\mathrm{O}^{+}$ and $24\pm 3$~ppb per ion for $^{134}\textrm{Ba}^{2+}$.   

This confirms for low intensities, particularly below $\approx 15$~ions per MR-TOF-MS cycle, ion-ion dependent shifts are smaller compared to the other systematic contributions.

To better understand the nature and behaviour of the observed ion-ion dependent shifts on the mass accuracy SIMION\cite{osti_912105_simion} simulations were performed at different intensities. An ion ensemble at mass $67$~u was simulated containing the five major ion species measured in the online spectrum. As shown in Fig.~\ref{Fig:A67TOF} mass $67$~u delivered at ISAC is dominated by $^{51}\textrm{V}^{16}\textrm{O}^{+}$ and $^{134}\textrm{B}^{2+}$ ions, with the remaining species $^{67}$Ga$^{+}$, $^{48}\textrm{T}^{19}\textrm{F}^{+}$ and $^{67}\textrm{Mn}^{+}$ making up less then $2\%$ of the total beam composition. 
To ensure sufficient statistics for fitting each simulation contained $1500$ ions. The SIMION build-in charge weighting factor was then adjusted for each species to result in an total ion intensity of $50$ and $100$ simultaneously flying ions while also ensuring the relative composition matches the delivered beam from ISAC. The simulated spectra were fitted and the relative movement of the individual peaks in comparison to $^{67}\textrm{Ga}^{+}$ was assessed. 

In simulation, the $^{51}\textrm{V}^{16}\textrm{O}^{+}$ and $^{134}\textrm{Ba}^{2+}$ peaks shift by about $15$~ppm per ion and $20$~ppm per ion, respectively. This is in excellent agreement with the experimentally observed shifts.   
Investigating the behaviour of the individual species further shows a rather different behaviour between the dominant ion species in comparison to the weaker ones, such as $^{67}\textrm{Mn}^{+}$ and $^{67}\textrm{Ga}^{+}$. The weaker species all seem to be pushed in the same direction by the rather concentrated charge density in the middle of the ion ensemble ($^{51}\textrm{V}^{16}\textrm{O}^{+}$ and $^{134}\textrm{B}^{2+}$). As a consequence, a reduced systematic shift is obtained for the mass of $^{67}\textrm{Mn}^{+}$ of only $4$~ppb per ion in simulation. However, given the limited statistics of the online measurement, this effect could not be investigated experimentally. This indicates that the effects of ion-ion interaction on the mass accuracy are in fact subtle and need to be assessed on a case by case basis. 

Our experimental ion-ion contribution of $16\pm2$~ppb per ion for singly charged ions should be seen as an upper limit systematic effect. Nonetheless, the uncertainty of the final mass values was increased according to the upper limit of the shift, which was combined in quadrature with the other systematic and statistical uncertainties. For the measurements of $^{66}$Mn, $^{67}$Mn and $^{68}$Mn, measured at $18$, $34$ and $19$ ions per cycle, relative uncertainties of $2.9$, $5.4$ and $3 \times 10^{-7}$ were added to obtain the final uncertainty reported in Tab. \ref{Tab:Masses}.         

\subsection{Final Mass Results}

\begin{figure}[t]
	\includegraphics[width=0.9\linewidth]{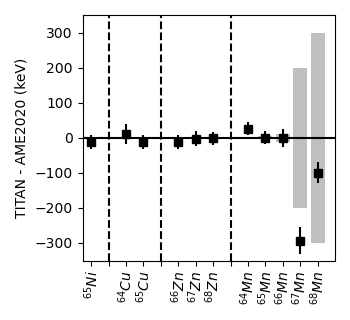}
	\caption{
		(color online)
		Mass differences of $^{65}\textrm{Ni}$, $^{64-65}\textrm{Cu}$, $^{66-68}\textrm{Zn}$, and $^{64-68}\textrm{Mn}$ between AME2020 and the present MR-TOF-MS measurements.
		The gray bands represent the $1\sigma$ error bars on the AME2020 value.
		The AME2020 values for $^{66,67}\textrm{Mn}$ are extrapolations.
    }
	\label{Fig:AllMnFitResults}
\end{figure}

The results of our measurements are summarized in Tab. \ref{Tab:Masses}, and the differences when compared to the AME2020 are shown in Fig. \ref{Fig:AllMnFitResults}.
The individual contributions to the final uncertainty were added in quadrature. 
The mass values of $^{65}\mathrm{Ni}$, $^{64,65}\mathrm{Cu}$, and $^{66-68}\mathrm{Zn}$ are well known and our measurements are in good agreement, which is an important cross check for our MR-TOF-MS results.
One additional (potential) source of systematic error in the present measurements would be long-lived isomeric states in the odd-odd $\mathrm{Mn}$ nuclides that may have been delivered simultaneously with the ground state.
In $^{64,66}\mathrm{Mn}$, the half-lives of any long-lived isomeric states have been measured to be below $1$\,ms \cite{PhysRevC.84.061305}, and thus would not survive the ion preparation sequence prior to being injected into the MR-TOF-MS.
No isomeric states have been observed in $^{68}\mathrm{Mn}$, however, if one were to exist, we believe that, based on systematics of the Mn isotopic chain, the half-life of this state should also be short enough that it would not impact our measurement.
In the cases of $^{64-66}\mathrm{Mn}$ the AME2020 values are dominated by the measurements from \cite{PhysRevC.86.014325}.
The present measurements are in good agreement, with the largest deviation of only $1\sigma$ at $^{64}\mathrm{Mn}$. The mass values obtained for $^{67}\mathrm{Mn}$ and $^{68}\mathrm{Mn}$ are the first precision measurements of these isotopes, and shows increased binding relative to the AME2020, as seen in Fig. \ref{Fig:AllMnFitResults}.

\begin{table}
\caption{Mass excesses as measured by the MR-TOF-MS. The reference isotope for each measurement is indicated in the table. The uncertainties on the TITAN mass excesses includes the systematic effects. Also included are the literature results from AME2020 \cite{Wang_2021}. Uncertainties marked with a \# are extrapolated values in the AME2020.}
\begin{tabular*}{\linewidth}{@{\extracolsep{\fill} } ccc}
	\hline
	Isotope(Reference) & $\mathrm{ME}_{\mathrm{TITAN}}$ (keV) & $\mathrm{ME}_{\mathrm{AME2020}}$ (keV) \\
	\hline \hline
	$^{65}\mathrm{Ni}(^{65}\mathrm{Ga})$ & $-65145\,(21)$ & $-65125.8\,\,(0.5)$ \\
	\hline
	$^{64}\mathrm{Cu}(^{64}\mathrm{Ni})$ & $-65413\,(28)$ & $-65424.4\,\,(0.4)$ \\
	$^{65}\mathrm{Cu}(^{65}\mathrm{Ga})$ & $-67276\,(19)$ & $-67263.7\,\,(0.6)$ \\
	\hline
	$^{66}\mathrm{Zn}(^{66}\mathrm{Ga})$ & $-68910\,(27)$ & $-68899.2\,\,(0.7)$ \\
	$^{67}\mathrm{Zn}(^{67}\mathrm{Ga})$ & $-67878\,(40)$ & $-67880.3\,\,(0.8)$ \\
	$^{68}\mathrm{Zn}(^{68}\mathrm{Ga})$ & $-70009\,(27)$ & $-70007.1\,\,(0.8)$ \\
	\hline
	$^{64}\mathrm{Mn}(^{64}\mathrm{Ni})$ & $-42963\,(19)$ & $-42989\phantom{.000}\,(4)$ \\
	$^{65}\mathrm{Mn}(^{65}\mathrm{Ga})$ & $-40967\,(19)$ & $-40967\phantom{.000}\,(4)$ \\
	$^{66}\mathrm{Mn}(^{66}\mathrm{Ga})$ & $-36750\,(26)$ & $-36750\phantom{.00}\,(11)$ \\
	$^{67}\mathrm{Mn}(^{67}\mathrm{Ga})$ & $-33874\,(39)$ & $-33580\,(200\#)$ \\
	$^{68}\mathrm{Mn}(^{68}\mathrm{Ga})$ & $-29020\,(30)$ & $-28920\,(300\#)$ \\
	\hline
\end{tabular*}
\label{Tab:Masses}
\end{table}

% ----------------------------------------------------
% -------------------- Discussion --------------------
% ----------------------------------------------------

\section{Discussion}

\begin{figure}
	\includegraphics[width=1.0\linewidth]{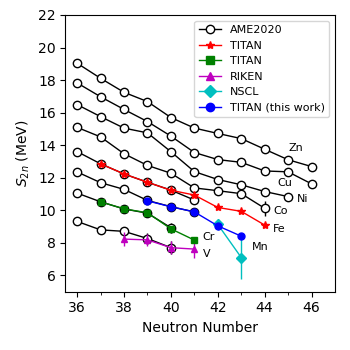}
	\caption{
		(color online)
		Two-neutron separation ($S_{\textrm{2n}}$) values in the vicinity of $N = 40$.
		Open circles are from the AME2020, excluding extrapolated values, while the filled circles in the $\textrm{Mn}$ chain are from the present work.
		The V, Cr, Mn, and Fe isotopic chains have been updated with the measurements from \cite{PhysRevLett.125.122501, PhysLettB.83.137288, PhysRevC.101.052801, PhysRevC.105.L041301} respectively.
	}
	\label{Fig:S2n}
\end{figure}

An important method to visualize shell structure and collectivity in the mass surface is the two-neutron separation energy $S_{\mathrm{2n}}$, which is a measure of the binding energy of the valence neutrons.
The two-neutron separation energy is defined as

\begin{equation}
\begin{split}
S_{\mathrm{2n}} &= B(Z,N) - B(Z,N-2) \\
\phantom{\,} &= -M(Z,N) + M(Z,N-2) + 2m_n,
\end{split}
\end{equation}

\noindent
where $B(Z,N)$ is the binding energy, $M(Z,N)$ is the atomic mass, and $m_n$ is the neutron mass.
Fig. \ref{Fig:S2n} shows $S_{\textrm{2n}}$ values in the vicinity of $N = 40$. As observed by other mass measurements in this region \cite{PhysRevLett.125.122501, PhysLettB.83.137288, PhysRevC.101.052801, PhysRevC.105.L041301}, there is no sharp drop in $S_{\textrm{2n}}$ across $N = 40$ indicating that no strong neutron subshell closure exists here. There is a strong opening that appears between Cr and Mn at $N = 41$, which may be a result of $^{65}$Cr being the expected center summit of the N=40 island of inversion \cite{PhysLettB.83.137288}.

\begin{figure}
	\includegraphics[width=\linewidth]{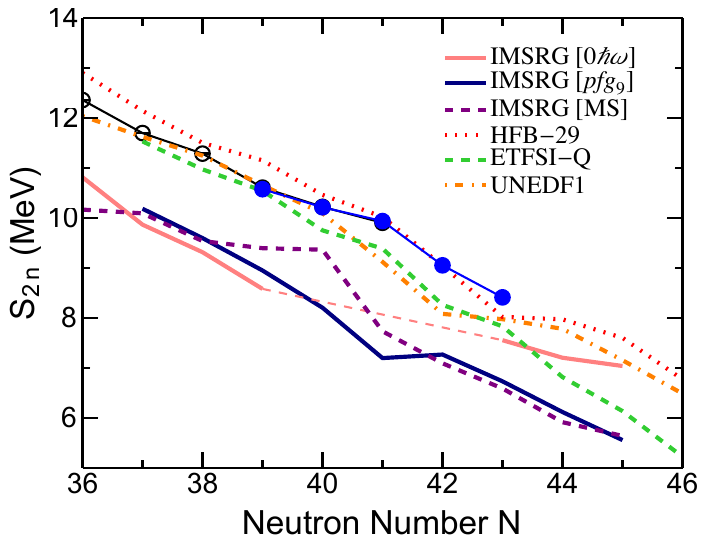}
	\caption{
		(color online)
		Comparison experimental and theoretical $S_{\mathrm{2n}}$ values in the Mn isotopic chain.
		Open points are the AME2016, filled points are the present measurements, the red, dotted line is the HFB-29, the green, dashed line is the ERFSI-Q, the orange, dot-dashed line is the UNEDF1, in addition to three different VS-IMSRG calculations described in the text.
	}
	\label{Fig:S2nTheory}
\end{figure}

\begin{figure}
	\includegraphics[width=\linewidth]{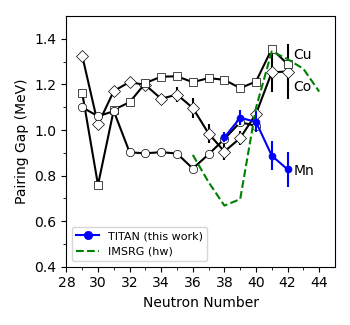}
	\caption{
		(color online)
		The empirical neutron pairing gaps from AME2020 as a function of neutron number for the odd-$Z$ isotopic chains of Cu, Co, and Mn (open symbols) and the present work for Mn (blue circles). Also shown is the prediction from the IMSRG [MS] calculation (green dashed line).
	}
	\label{FIG:neutronPairingGap}
\end{figure}

In Fig. \ref{Fig:S2nTheory}, we show the Mn $S_{\mathrm{2n}}$ values, along with several different theoretical calculations.
In general, the energy density functional results with the HFB-29 \cite{GORIELY201568}, ESTFI-Q \cite{PEARSON1996455}, and UNEDF1 \cite{PhysRevC.85.024304} functionals, agree quite well with experimental values below $N = 40$. After $N = 40$, however, the ETSFI-Q and UNEDF1 results exhibit too little binding with increasing neutron number, while, in contrast, the HFB-29 agrees remarkably well over the entire range.
Neither ETFSI-Q nor HFB-29 gives a pronounced subshell closure at $N = 40$, while the UNEDF1 calculation shows a clear change in slope at $N = 40$, indicating a subshell closure.
In order to further elucidate the onset of collectivity near $N = 40$ through the mass surface we also compare to predictions from state-of-the-art \textit{ab initio} calculations from the valence-space formulation of the in-medium similarity renormalization group (VS-IMSRG)~\cite{Stro19ARNPS}, which enable calculations of all open-shell nuclei accessible with the standard shell model approach.
These calculations are performed with two- ($NN$) and three-nucleon ($3N$) forces based on chiral effective field theory~\cite{Epel09RMP}.
In particular we use the $1.8/2.0$(EM) interaction of Refs~\cite{Hebe11fits,Simo17SatFinNuc}, which is fit only to properties of few-body systems, but nevertheless accurately reproduces ground-state energies to the tin region~\cite{PhysRevLett.126.022501,Morr17Tin,PhysRevLett.124.092502}, including physics in between calcium and nickel~\cite{XU2019,LEIS2021,PhysLettB.83.137288,PhysRevLett.125.122501,PhysRevC.97.014309,PhysRevC.105.L041301}.
We use the imsrg++ code~\cite{Stro17imsrg++} in the IMSRG(2) approximation where induced operators are truncated at the two-body level in the Magnus expansion~\cite{Morr15Magnus}. The ensemble normal ordering procedure of Ref.~\cite{Stro17ENO} captures the physics of 3N forces between valence particles. While we always use $pf$ orbits for the proton valence space, we explore several choices for the neutron space in order to best capture the physics across $N=40$: for [$0\hbar \omega$] we take neutron $pf$ orbits for $N<40$ and neutron $sdg$ orbits for $N>40$, and for [$pfg_9$] we take the $p_{3/2}$, $p_{1/2}$, $f_{5/2}$, and $g_{9/2}$ neutron orbits. In addition for IMSRG [MS], we use the newly proposed multi-shell VS-IMSRG technique \cite{PhysRevC.102.034320} with the $pfg_9$ neutron valence space, removing spurious center-of-mass components via the procedure outlined therein.

Both the [$0\hbar \omega$] and [$pfg_9$] calculations generally underbind the experimental $S_{\mathrm{2n}}$ values presented, noting that the [$pfg_9$] calculations evolve in a similar manner to the AME2020 extrapolated values, while still underbinding relative to the present measurements.
The new results do capture an increase in collectivity at $N = 40$, however, they still follow the trend of the [$pfg_9$] calculations, which is similar to that seen in the chromium isotopic chain~\cite{PhysRevLett.120.232501}, with an analogous underbinding beginning at $N = 36$.
As discussed there, the underbinding of the IMSRG calculations in this region is likely due, at least in part, to the truncation of the many-body terms at the IMSRG(2) level, where work is currently in progress to estimate the next many-body order~\cite{HE2024} on the $N=40$ island of inversion.

This is in juxtaposition to VS-IMSRG calculations in the proton and neutron shells $pf_{5/2}g_{9/2}$ for the Ni isotopes, which adequately predicts the behavior across $N=40$ \cite{PhysRevC.97.014309}.
As has been noted in other shell-model calculations in this region, the $d_{5/2}$ orbital becomes important in the mid-shell $N = 40$ nuclei below $^{68}\mathrm{Ni}$, as particle-hole excitations across this diminished gap drive the onset of collectivity \cite{PhysRevC.89.024319,PhysRevC.82.054301}.
In contrast to the island of inversion near $N = 20$ where the ground state wave functions are dominated by $2p-2h$ excitations, the wave functions below $^{68}\mathrm{Ni}$ are dominated by $4p-4h$ excitations with $^{64}\mathrm{Cr}$ having the largest gain in correlation energy \cite{PhysRevC.82.054301}.
We also note that the ground-state of $^{68}\mathrm{Ni}$ is largely made up of $0p-0h$ and $2p-2h$ configurations, and the IMSRG calculations are able to reproduce the $S_{2n}$ values near $N=40$.
In the VS-IMSRG approach, the too-large gap at $N=40$ likely energetically inhibits $p-h$ excitations in the region, to begin with, and furthermore collectivity in general is a challenge to capture in \textit{ab initio} approaches, unless for example, novel collective references are employed~\cite{PhysRevLett.124.232501}.
In addition there is a non-negligible dependence on the assumed filling of neutron orbits for inclusion of 3N forces between valence particles, which may also influence the differences observed between [$pfg_{9}$] and [MS] calculations.

To better understand the onset of collectivity in the Mn isotopes across $N = 40$, we determined the neutron pairing gap from the present mass measurements.

The neutron pairing gap can be approximated as
\begin{equation}
\begin{split}
	P_\mathrm{n} (Z, N) =& \frac{(-1)^{N+1}}{4} ( S_\mathrm{n} (Z,N+1) - 2S_\mathrm{n} (Z,N)\\
	\phantom{\,} & + S_\mathrm{n} (Z,N-1) ),
\end{split}
\end{equation}

\noindent
where $S_\mathrm{n} (Z,N)$ is the one-neutron separation energy.
The neutron pairing gap for the Mn isotopic chain is shown in Fig. \ref{FIG:neutronPairingGap}. 

There is an increase in the pairing gap from $N = 38$ through to $N = 40$ and then decreasing beyond $N = 40$, which indicates an increase in collectivity.
This increase in the pairing gap is most likely related to neutrons occupying higher lying orbitals, with a corresponding decrease in the extent of correlation when neutrons states are populated past $N = 40$.

In Fig. \ref{FIG:neutronPairingGap} we also compare the Mn neutron pairing gap to the odd-$Z$ isotopic chains of Cu and Co.
In the case of Cu ($Z = 29$), the neutron pairing gap is relatively constant, which most likely indicates a relatively constant gap between single particle energies for the different orbitals.
In contrast, the Co and Mn isotopic chains exhibit opposite behavior, where the Cu chain dips, reaching a minimum at $N=38$, while in the Mn chain peaks, reaching a maximum at $N=39$. The increase in the neutron pairing gap indicates that neutrons are beginning to occupy higher orbitals from $N=38$. This was also observed by ISOLTRAP \cite{PhysRevC.86.014325} in their measurement of isotopes in the same region.
The IMSRG [MS] calculation follows the same behavior as in the Co chain, also reaching a minimum at $N=38$.
While it is difficult to separate possible shell effects from collective behavior, the opposite behavior of the data and 
\textit{ab initio} calculations in the Mn chain point towards an increase in collectivity approaching $N=40$ which are not fully captured by the calculations.

% ----------------------------------------------------
% --------------------- Summary ----------------------
% ----------------------------------------------------

\section{Summary}

We have measured the neutron-rich $^{64-68}\mathrm{Mn}$ isotopes with the multiple-reflection time-of-flight mass spectrometer device at TITAN and for the first time investigated effects of ion-ion interactions on the mass accuracy of an MR-TOF-MS with short-lived ions. The observed behavior is well matched by simulation allowing the systematic effect to be characterized for high-rate measurements.
These final measurements represent the first precision results of $^{67}\mathrm{Mn}$ and $^{68}\textrm{Mn}$, extending the measured masses well past the $N = 40$ subshell closure to $N = 43$.
The results were compared to state-of-the-art \textit{ab initio} calculations, where large discrepancies, peaking at $N = 41$, to the experimental data were found.
These discrepancies point to the importance of multiparticle-multihole excitations in this region, which are not fully captured in the \textit{ab initio} calculations.
The new masses expand the mass surface and provide an important testing ground for the development of next generation \textit{ab initio} models in the theoretically challenging region around $N=40$. 

\section{Acknowledgements}

The authors would like to thank M. Good for his continual support and J. Men\'{e}ndez for his valuable discussions.
We are grateful to S. R. Stroberg for the imsrg++ code used to perform the calculations.
This work was supported by Canadian agencies NSERC and CFI, the U.S. Department of Energy by Lawrence Livermore National Laboratory under Contract No. DE- AC52-07NA27344, Brazil's CNPq (grant 249121/2013-1), the Canada-UK Foundation, German institution BMBF (grants 05P15RGFN1 and 05P12RGFN8) and by the JLU and GSI under the JLU-GSI strategic Helmholtz partnership agreement.
J. Lassen, R. Li and M. Mostamand acnowledge support from TRIUMF.
M. Mostamand further acknowledges funding through the University of Manitoba graduate fellowship. For the purpose of open access, the author has applied a Creative Commons Attribution (CC BY) licence to any Author Accepted Manuscript version arising from this submission.

\bibliography{bib_final}

\end{document}